\documentclass[12pt]{iopart}
\usepackage{epsfig}
\begin{document}
\title{Chiral structures of lander molecules on Cu(100)}
\author{J.~Kuntze, X.~Ge, R. Berndt}
\address{Institut f\"ur Experimentelle und Angewandte Physik, Christian-Albrechts-Universit\"at zu Kiel, Olshausenstr. 40, D-24098 Kiel, Germany}
\date{\today}
\maketitle
\begin{abstract}
Supramolecular assemblies of lander molecules (C$_{90}$H$_{98}$) on Cu(100) are investigated with low-temperature scanning tunneling microscopy. The energetically most favourable conformation of the adsorbed molecule is found to exist in two mirror symmetric enantiomers or conformers. At low coverage, the molecules align in enantiomerically pure chains along the chiral directions $[01\bar{2}],[02\bar{1}],[012]$ and $[021]$. The arrangement is proposed to be mainly governed by intermolecular van-der-Waals interaction. At higher coverages, the molecular chains arrange into chiral domains, for which a structural model is presented.   
\end{abstract}
\pacs{68.43.-h, 68.37.Ef, 33.15.Bh}

\vspace{1ex}

Chiral surfaces are of great interest due to their importance in enantioselective catalysis, e.g. in pharmaceutical industry. Using chiral catalysts, the yield of products with the desired handedness can be drastically increased \cite{izumi_83a}. Apart from this industrial perspective, the study of chiral molecules is fascinating from a fundamental point of view. Since the recent determination of absolute chirality of adsorbed molecules by scanning tunneling microscopy (STM) \cite{lopinski_98a}, many interesting phenomena involving chiral structures have been observed over the last few years. E.g., chiral supramolecular assemblies of nitronaphtalene clusters were found \cite{boehringer_99a} that could be manipulated using the STM tip \cite{boehringer_99b}, chiral recognition at the single molecule level has been reported \cite{boehringer_99b,boehringer_00b,kuehnle_02a} and the chiral restructuring of surfaces due to adsorption of hydrocarbon molecules has been observed \cite{schunack_01a,schunack_01b}. 

Chirality can be bestowed onto a surface in a number of ways. One is to cut the surface along a high Miller-index direction leading to kinked surfaces that can be chiral \cite{sholl_01a}, another is the formation of supramolecular assemblies of chiral molecules \cite{lorenzo_00a}. Both methods would lead to overall, global chirality. {\it Local} chirality can be observed when a non-chiral molecule creates a chiral motif upon adsorption and eventually assembles into supramolecular geometries which are chiral themselves. Here we report on such a case, where the so-called lander molecule (C$_{90}$H$_{98}$) \cite{kuntze_02a,rosei_02a,rosei_03a}, a prototype of a short molecular wire, is adsorbed on Cu(100). The molecule is found to adopt two mirror symmetric conformations which assemble into enantiomerically pure domains with chiral symmetry. Macroscopically, the surface is still racemic owing to the presence of two mirror symmetric molecular conformations.

The experiments were performed with a home-built low-temperature STM \cite{kliewer_thesis} in ultra-high vacuum. Base
pressures are in the low 10$^{-10}$ mbar range and below 10$^{-11}$
mbar for the preparation and STM-chambers, respectively. Atomically clean Cu(001) surfaces were
prepared by standard procedures, namely repeated cycles of sputter
cleaning (1 keV Ne ions) at room temperature followed by subsequent
annealing to approx.$\,$700 K for 15 min. The molecules were
sublimated from a Knudsen cell at an evaporation rate of less than
0.005 Monolayers/min, as monitored by a quartz crystal
microbalance. A monolayer is defined here as one molecule per ($1\times1$)-cell of the Cu(100) substrate. The sample was held at room temperature during
deposition and was transferred to the STM and cooled down to 4.6 K
over a period of typically 12 hours prior to imaging. No postannealing was done after molecule deposition.

Electrochemically etched tungsten tips were prepared {\it in vacuo}
by electron bombardement and Ne$^+$ sputtering before insertion in
the STM. During image acquisition, the tips were further treated by field
emission and controlled tip-sample contacts with the tip
positioned over bare copper terraces to avoid contamination with molecules or
molecular fragments. Owing to this tip sharpening procedure, we
assume that the tip apex is comprised of Cu atoms.

\begin{figure}
\centerline{\epsfig{figure=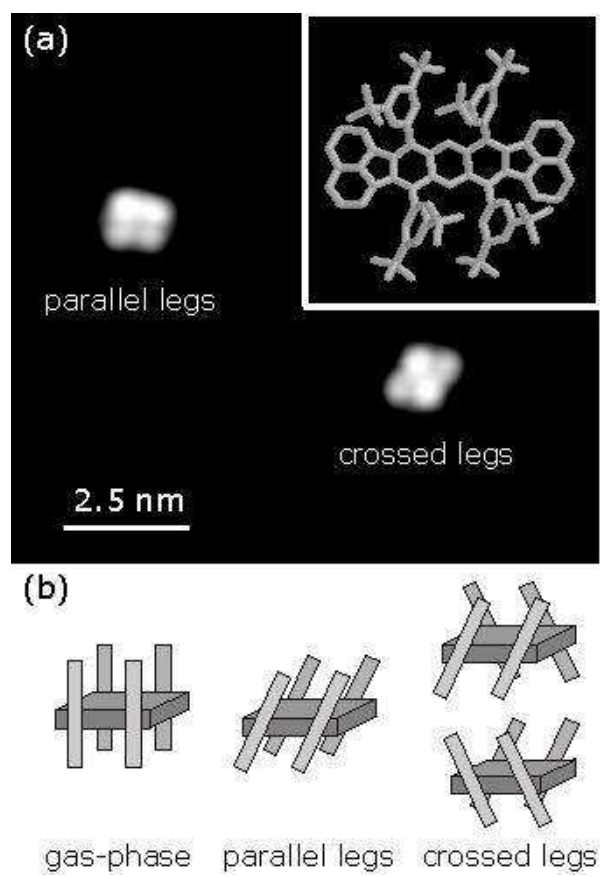,width=0.5\linewidth,clip=}}
\vspace{1ex}
\caption{\label{fig1} Two isolated lander molecules on Cu(100) in different conformations.
Image size (9 nm)$^2$, tunneling current I= 50 pA, sample voltage U= 2 V.  The left molecule is in the parallel-legs conformation, the right one in the crossed-legs conformation. A model of the lander in the crossed-legs conformation is shown as an inset (with hydrogen atoms omitted for clarity). Below the image, the conformations are schematically illustrated. The molecule is sketched as a board with four legs, the DTP groups, attached to it. }
\end{figure}

Figure \ref{fig1}a shows two isolated lander molecules on a terrace of Cu(100). Each molecule is imaged as four bright lobes, which are due to tunneling through the DTP {\it di-tert-butyl-phenyl})-spacers attached to the central polyaromatic board (cf.\ to the inset of Fig.\,\ref{fig1}a for a ball-and-stick model of the molecule). The DTP-substituents are designed to lift the central polyaromatic board above the substrate. As a result, it does not contribute significantly to the tunnelling current. Two distinct molecular conformations are observed, which are schematically depicted in Fig.\,\ref{fig1}b. The DTP-legs  rotate around the $\sigma$-bonds attaching them to the molecular backbone. If all legs tilt in the same direction, a square shape conformation results (parallel-legs). If the two legs on one side of the molecular board tilt in the opposite direction than the legs on the other side, a rhombic shape results (crossed-legs). The reason why the four protrusions corresponding to the DTP-spacers appear with two distinct heights is the interaction of the legs upon rotation, leading to more effective tunneling channels for two of the legs, which results in one bright and one dim protrusion on each side of the molecular board \cite{kuntze_02a}. From comparison of experimental images with elastic scattering quantum chemistry (ESQC) image simulations, a strong perturbation of molecular bonds as compared to the free molecule was inferred \cite{kuntze_02a}. The central board was found to be lowered to a height of 0.37 nm above the surface, which is achieved via a bending of the $\sigma$-bonds. 

\begin{figure}
\centerline{\epsfig{figure=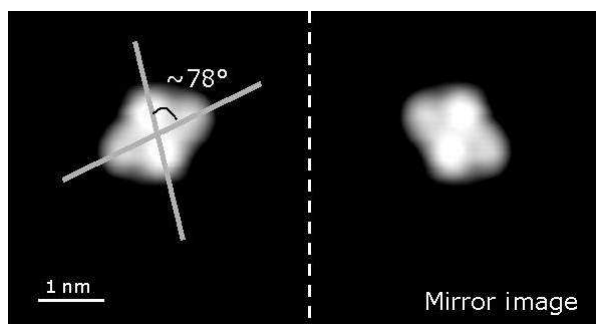,width=0.5\linewidth,clip=}}
\vspace{1ex}
\caption{\label{fig2} Close view of the cross-legged molecule of Fig.\protect\ref{fig1}. The lines indicate the direction of molecular axes. On the right side, the mathematical mirror image of the molecule is displayed.}
\end{figure}

Of the two principal conformations, the crossed-legs version is energetically preferred and thus far more abundant experimentally. Figure \ref{fig1}b shows that there are two ways to cross the legs on both sides of the molecule, namely clockwise or anticlockwise rotation of the legs on on side of the central board (when viewed from the side), each leading to a nearly rhombic image. One can define a long and short axis of the molecule (cf.\,Fig.\,\ref{fig2}), connecting the dim and bright protrusions in the STM image, respectively. Close inspection of the cross-legged molecule in Fig.\,\ref{fig2} reveals that the short and long axis are not perpendicular to each other, as they would be in a rhombus. The molecular outline is rather a parallelogram and thus chiral, since mirror images of this conformation cannot be superimposed onto each other by translation or rotation.  This chirality of the individual molecules has implications for molecular assemblies.

\begin{figure}
\centerline{\epsfig{figure=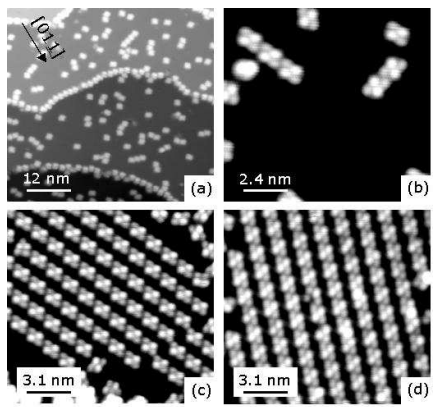,width=0.5\linewidth,clip=}}
\vspace{1ex}
\caption{\label{fig3} Assembly of (a-b) short chains and (c-d) extended supramolecular domains. The orientation of all images is identical, as indicated by the arrow in (a). Only molecules of identical symmetry are observed to align. The domains in (c-d) are mirror symmetric with respect to the $[01\bar{1}]$ direction. Two more equivalent domain orientations which are symmetric with respect to the [011] direction, are not shown here. } 
\end{figure}

At increasing coverage, the molecules arrange themselves in short chains along $[01\bar{2}],[02\bar{1}],[012]$ and $[021]$ directions (Fig.\,\ref{fig3}a,b). The crystallographic $\langle 110 \rangle$ directions as determined from low-energy electron diffraction (LEED) and extended straight atomic steps produced by tip-sample contact are indicated in the figure. The molecular chains are oriented at an angle of approx.\,18$^\circ$ with respect to these high-symmetry directions. Care was taken to exclude image distortion by thermal drift or piezo creep. Given the residual uncertainties in scanner calibration we estimate the uncertainty of the angles to $\pm 2^\circ$. 

Interestingly, only molecules of like symmetry align in the same chain. This finding is consistent with previous observations  for other racemic mixtures of chiral molecules (either intrinsically 3D-chiral or 2D-chiral after adsorption) \cite{barlow_03a,defeyter_03a}. In most cases, chiral discrimination can be traced back to functional groups interacting via hydrogen bonding, e.g., hydroxy groups \cite{lorenzo_00a,barth_00a} or carboxylic groups \cite{kuehnle_02a}. The lander molecules, however, do not exhibit such groups and can be expected to interact predominantly via van-der-Waals forces. Since there are small partial charges associated with the CH$_3$ groups on the DTP-legs, weak local dipole moments exist on the legs, but they are oriented mainly perpendicular to the plane of the molecular board and will contribute only little to the in-plane molecular arrangement.  A similar case of spontaneaous resolution into conglomerates for van-der-Waals dominated interaction was reported for Naphto[2,3-{\it a}]pyrene \cite{france_03a}. 

The  molecular chain directions do not correspond to symmetry directions of the sample, which makes the assembly itself chiral, without taking into account the chirality of the individual molecules. At higher coverages the molecules assemble into domains along the same directions  as the short chains. Figures \ref{fig3}c,d show two such domains which are mirror symmetric with respect to the $[01\bar{1}]$ direction (cf.\ arrow in Fig.\,\ref{fig3}a). One readily verifies that also the molecular units are mirror images of each other. For the domains, the distance between molecular centers in adjacent rows is $(1.63\pm 0.1)$ nm and the distance between molecules in a row is $(1.56\pm 0.1)$ nm, with an angle of approximately $(84\pm 2)^\circ$ between these two directions. The angle between molecular chains in mirror symmetric domains as in Fig.\,3c,d is $(36\pm 2)^\circ$. The other two equivalent domains symmetric with respect to the [011] direction are not shown here. We note that extended disordered regions remain between the domains, predominantly with shorter chains and few isolated molecules which have neither assembled into chains nor domains. The size of ordered domains is of the order of 15 nm in diameter. We suggest that by fine-tuning the growth conditions as deposition rate and substrate temperature, the domain size may be increased. 

\begin{figure}
\centerline{\epsfig{figure=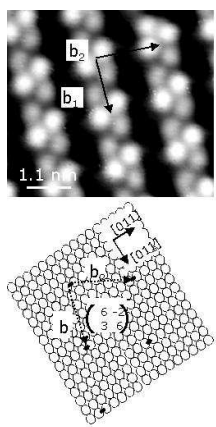,width=0.5\linewidth,clip=}}
\vspace{1ex}
\caption{\label{fig4} Close view of the structure within a domain and corresponding overlayer registry. The model is rotated to match the crystallographic orientation of the experimental image. The overlayer is given in matrix notation, with the unit vectors ${\bf b_1,b_2}$ defined with respect to the substrate unit vectors ${\bf a_1}\| [01\bar{1}], {\bf a_2}\| [011]$.}
\end{figure}

Equipped with the data given above, we have sought to model the molecular arrangement. Figure \ref{fig4} shows a closeup of  a molecular domain with the unit vectors of the overlayer and the corresponding model. The unit vector ${\bf b_1}$ along the direction of molecular rows is oriented along $[01\bar{2}]$ for the domain orientation shown, and the unit vector across the rows ${\bf b_2}$ is oriented along [031]. In terms of the substrate unit vectors ${\bf a_1}= a_0/2 [01\bar{1}]$ and ${\bf a_2}= a_0/2 [011]$, $a_0=0.362$ nm being the Cu lattice constant, the overlayer can be given in matrix notation as (6 -2, 3 6). The lengths of ${\bf b_1}$ and ${\bf b_2}$ are 1.62 and 1.72 nm, respectively, and the angle enclosed by the vectors is $82^\circ$. The angle between {\bf $b_1$} vectors in mirror symmetric domains is 36.8$^\circ$ (each is 18.4$^\circ$ with respect to the symmetry directions $\langle 110 \rangle$). This agrees well with the experimental values within the estimated uncertainties. As the molcular adsorption site cannot be determined from our STM images, the unit vectors in Fig.\ref{fig4} are arbitrarily chosen to connect bridge sites. The matrix notation for the mirror symmetric domain (cf.\ Fig.\,\ref{fig3}c) is (6  2, 3 -6), for the other two domains we find (-2 6, 6 3) and (2 6, -6 3). The overlayer unit vectors are always chosen according to the model in Fig.\,\ref{fig4}, with ${\bf b_1}$ pointing along the chain direction and ${\bf b_2}$ pointing across the chains, with an angle of less than 90$^\circ$ between the unit vectors. The overlayer unit vectors are always expressed in terms of the substrate unit vectors ${\bf a_1,a_2}$ given above.  

The question why the molecules order along the given directions is still open. Formation of directed molecular chains is a natural consequence of a fixed orientation of a single molecule with respect to the substrate and a preferred intermolecular alignment. Electrostatic interactions should favour a close packing of the molecules to increase the attractive forces. The configuration which maximises molecular interactions will then be energetically preferred and result in a specific intermolecular arrangement. This may also explain the spontaneous resolution into homochiral phases, provided that homochiral interactions dominate. Such a scenario was indeed observed previously \cite{boehringer_00b}.

In summary, we have observed chiral arrangements of lander molecules on Cu(100). At low coverage, the molecules assemble in enantiomerically pure molecular rows along the $[01\bar{2}],[02\bar{1}],[012]$ and $[021]$ directions. These rows arrange into enantiomerically pure domains at higher coverage, for which an overlayer model is presented. The overlayer geometry is also chiral by itself but due to the presence of mirror symmetric molecules the surface is racemic on a macroscopic scale. The molecular interactions responsible for the chiral arrangements are proposed to be mainly due to van der Waals forces and steric effects, since no substantial partial charges or strong hydrogen bonds exist for these molecules.

{\bf Acknowledgement} We acknowledge financial support from the EU
IST-FET project "Bottom up Nanomachines" and from the DFG under the EUROCORES project "Self organization of nanostructures" (SONS). We thank Hao Tang and Andr{\'e} Gourdon for discussions.



\newpage

\newpage
\newpage

\end{document}